\documentclass[twoside]{elsart}
\usepackage{epsfig}
\usepackage{psfig,wrapfig,amsmath,amstext}

\begin{document}
{
\null \hfill \begin{minipage}{5cm} Freiburg-THEP-06/02\\
In memory of Alfred Hill
\end{minipage}
}
\begin{center}
\title{The minimal non-minimal standard model}
\author{J. J. van der Bij}
\address{Institut f\"ur Physik, Universit\"at Freiburg,\\
  Hermann-Herder-Str. 3, 79104 Freiburg
i. Br., Germany}
\end{center}
{\bf Abstract}\\
In this letter I discuss a class of extensions of the standard model
that have a minimal number of possible parameters, but can in principle
 explain dark matter and inflation. It is pointed out that the so-called 
new minimal standard model contains a large number of parameters
that can be put to zero, without affecting the renormalizability
of the model. 
With the extra restrictions one might call it 
the minimal (new) non minimal standard model (MNMSM).
A few hidden discrete variables are present. It is argued that the
inflaton should be higher-dimensional. Experimental
consequences for the LHC and the ILC are discussed.

With the latest developments from high energy colliders like LEP
and the Tevatron the standard model (SM) has been established up to
the loop level. Precision measurements leave only very little
space for extensions, as these tend to spoil the agreement with experiment
due to a variety of effects, one of the most important of which is
the appearance of flavor-changing neutral currents. Even the most
 popular extension,
namely the minimal supersymmetric extension of the SM has to finely
tune a number of parameters. This leaves only one type of extensions that are
safe, namely the
singlet extensions. Experimentally  right handed
neutrinos appear to exist. Since these are singlets a natural extension of the SM
is the existence of singlet scalars too [1-9]. These will only have a very limited effect
on radiative corrections, since they appear only in two-loop calculations [10,11].\\

The effects of singlets appear in two forms, one is the mixing with the SM
Higgs, the other is the possibility of invisible decay. In contrast to charged fields
these effects can be separated. It is actually possible to have a Higgs model
that has only Higgs-mixing. If one starts with an interaction of the form
$H \Phi^{\dagger}\Phi $, where H is the new singlet Higgs field and $\Phi$ the SM
Higgs field, no interaction of the form $H^3$ or $H^4$ is generated with an infinite
coefficient \cite{hill}. This means that one can indeed leave these interactions out of the Lagrangian
without violating renormalizability. This is similar to the non-renormalization theorem
in supersymmetry that says that the superpotential does not get renormalized.
However, here it only works with singlet extensions. As far as the counting of parameters
is concerned this is the most minimal extension of the SM, having
only two extra parameters.
It has been called the most-minimal non-minimal standard model (MMNMSM).
The model descibes a SM Higgs boson, that is split into two Higgs
bosons, that each have the SM branching ratios, but that are produced with
a reduced rate. Actually combining the LEP2 data
slight excesses in events are present at $98\,GeV$ and $115\,GeV$,
that however are still compatible with background \cite{LEPHiggs}. Although such an
excess is clearly incompatible with a SM Higgs signal, it can be easily explained
  by a fractional Higgs boson. This model can be straightforwardly extended 
with more Higgs singlets [8,9],
so as to produce an arbitray line shape for the Higgs boson, satisfying
the K\"all\'en-Lehmann representation for the Higgs field.
$D_{Higgs}(k^2)= \int ds\, \rho(s)/(k^2 + \rho(s) -i\epsilon)$, whereby
one has $\int \rho(s)\, ds = 1$, while otherwise the theory is not renormalizable
and would lead to infinite effects for instance on the LEP precision variables \cite{akhoury}.
Actually the $H$ field does not even have to be four dimensional,
 since the $H \Phi^{\dagger} \Phi$ coupling constant has a positive mass dimension
as long as the $H$ field lives in $d < 6$ dimensions. The critical case
$d=6$ leads actually to a tree level renormalization group running of the   
$\Phi^4$ and $H \Phi^{\dagger} \Phi$ couplings [13-15].\\

To show how these models behave I will explicitely show the Lagrangian of the
simplest model [1] and the models with flat noncompact open dimensions for the
singlet field $H$. These can be studied analytically.
The model [1] was recently written down again in a different
notation \cite{slavnov}.
The model of \cite{krasnikov}
introduced the five dimensional continuous mass  Higgs boson.
It is not very specific about the Higgs boson self-interactions, however 
appears to be nonrenormalizable,
since it allows for a $H^4$ coupling even in the higher dimensions.
The simplest model is the Hill model: 
\begin{equation}
L = -\frac{1}{2}(D_{\mu} \Phi)^{\dagger}(D_{\mu} \Phi) 
-\frac {1}{2}(\partial_{\mu} H)^2 - \frac {\lambda_0}{8}
(\Phi^{\dagger} \Phi -f_0^2)^2  -
\frac {\lambda_1}{8}(2f_1 H-\Phi^{\dagger}\Phi)^2 
\end{equation}
Working in the unitary gauge one writes $\Phi^{\dagger}=(\sigma,0)$,
where the $\sigma$-field is the physical SM Higgs field.
Both the SM Higgs field $\sigma$ and the Hill field $H$ receive vacuum expectation
values and one ends up with a two-by-two mass matrix to diagonalize, thereby
ending with two masses $m_-$ and $m_+$ and a mixing angle $\alpha$. There are two
equivalent ways to describe this situation. One is to say that one has two Higgs
fields with reduced couplings g to SM particles:
\begin{equation}
g_-= g_{SM} \cos(\alpha), \qquad g_+= g_{SM} \sin(\alpha)
\end{equation} 
The other way, which has some practical advantages is not to diagonalize
the propagator, but simply keep the $\sigma - \sigma$ propagator explicitely.
One can ignore the $H-\sigma$ and $H-H$ propagators, since the $H$ field does not
couple to ordinary matter. One simply replaces in all experimental cross section
calculations the SM propagator by:
\begin{equation}
D_{\sigma\sigma} (k^2) = \cos^2(\alpha)/(k^2 + m_-^2) + \sin^2(\alpha)/(k^2 + m_+^2)
\end{equation}
The generalization to an arbitrary set of fields $H_k$ is straightforward, one 
simply replaces the singlet-doublet interaction term by:
\begin{equation}
L_{H \Phi}= - \sum \frac {\lambda_k}{8}(2f_k H_k-\Phi^{\dagger}\Phi)^2 
\end{equation}
For a finite number of fields $H_k$ no essentially new aspects appear.

The situation becomes more interesting, when one allows for an infinite number
of fields, for instance when one assumes the field $H$ to be moving in 
$d=4+\delta$ dimensions.
Normally speaking this would lead to a nonrenormalizable theory. However
since the only interaction is of the form
$H\Phi^{\dagger}\Phi$, which is superrenormalizable in four dimensions,
the theory stays renormalizable. An analysis of the power counting of divergences
shows that one can associate the canonical dimension $1+\delta/2$ to the
$H$-field. This means that the theory stays renormalizable as long as
$\delta \le 2$. 
When one assumes, that the extra dimensions are compact, for instance a torus,
with radius $R=L/2 \pi$,
one can apply the formalism above, simply by expanding the $d$-dimensional field
$H$ in terms of Fourier modes:
\begin{equation}
H (x) ={1 \over \sqrt{2} L^{\delta/2}} \sum_{\vec{k}} \phi_{\vec{k}}(x_\mu)
e^{i {2 \pi \over L} \vec{k} \vec{x} } \;\;\;\;\;\;\;\;\;\;\;\;
  H_{\vec{k}}      =   H_{-\vec{k}}^*        
\end{equation}

Here $x_\mu$ is a four-vector, $\vec{x}$ is $\delta$-dimensional
and the $\delta$ components of $\vec{k}$ take only integer values.

One then finds a Lagrangian of the form:
 \begin{equation}
  L=-\frac{1}{2}D_{\mu} \Phi^{\dagger} D_{\mu} \Phi 
 - {1 \over 2} \sum (\partial_{\mu} H_k )^2
 +{M^2 \over 4} \Phi^{\dagger}\Phi 
-{\lambda \over 8} (\Phi^{\dagger}\Phi)^2
 -\sum {m_k^2\over2} H_k^2 
 -{g \over 2} \Phi^{\dagger}\Phi  \sum H_k 
\end{equation}
The masses of the $H_k$ fields are given by $m_k^2=m_0^2+m^2\vec k^2$.
Here $\vec k$ is a $\delta$-dimensional vector and $m_0$ is a $d$-dimensional
mass term for the field $H$.  This term is necessary to insure stability
of the vacuum. The last term can be written as a so-called brane-bulk term
\begin{equation} 
 S=\int d^{4+\delta}x \prod_{i=1}^{\delta} \delta(x_{4+i}) H(x) \Phi^{\dagger}\Phi   
\end{equation}
After minimizing the potential one has to diagonalize
 an $\infty$-by-$\infty$ mass matrix, which
is possible but not very illuminating. It is actually sufficient to
look at the $\sigma-\sigma$ propagator, which becomes in the continuum limit:

\begin{equation}
D_{\sigma\sigma}(q^2)= \left[ q^2 +M^2 -g^2 v^2 {\Gamma (1-\delta/2) \over (4 \pi)^{\delta/2} }
(q^2+m_o^2)^{\delta-2 \over 2} \right]^{-1}
\end{equation}

In the limit $\delta=0$ one regains the propagator of the Hill model.
In the limit $\delta \rightarrow 2$ one notices that the propagator
has  a logarithmic singularity. This is because the coupling constant
becomes dimensionless and is running as a function of the renormalization scale.
In contrast to normal four-dimensional models the running appears
already at the tree level of the theory. What is important for phenomenology
is that the K\"all\'en-Lehmann density $\rho(s)$ extends to infinity,
asymptotically one has:
\begin{equation}
\rho(s) \sim s^{-3+\delta/2}\qquad s\rightarrow\infty
\end{equation}
In this sense one can also speak about a $H$ field with a fractional
dimension. The fact that the density extends to infinity has as a consequence
that these models behave a bit more like a heavy Higgs field in the
fits of the electroweak precision data. They are therefore not really
preferred by the electroweak fits. However since the mixing with the
higher dimension is controlled by a coupling constant, they are also not ruled
out by the precision data.\\

The other possibility with singlet Higgs fields is pure invisible decay [7].
This actually is very natural, when one introduces Higgs singlets $S_i$ as 
multiplets of a symmetry group, for instance $O(N)$. When the $O(N)$ symmetry group
stays unbroken this leads to an invisibly decaying Higgs boson through
the interaction $\Phi^{\dagger}\Phi S_i S_i$, after spontaneous breaking of the
SM gauge symmetry. Typically the invisible width will be larger than
the SM width by far, as this decay is not suppressed by the very small
Yukawa coupling to the quarks. In this case the Higgs is wide and decaying 
invisibly. This would make it rather difficult to detect at the LHC,
which explains the name stealthy Higgs model for this type of models. 
Excluding the "hidden" parameter $N$ this model has three extra parameters
beyond the SM and is therefore not as minimal as the pure mixing models.
Also a combination of mixing and invisible decay is simply possible,
when one breaks the $O(N)$ symmetry spontaneously [2-6].\\

To be more concrete let us discuss the Lagrangian of the
pure decay model,  containing the standard model
Higgs boson plus an O(N)-symmetric sigma model.
The Lagrangian density is the following:

\begin{equation} 
L_{Scalar} = L_{Higgs} + L_{S} + L_{Interaction}     
\end{equation}
\begin{equation}
L_{Higgs}  = 
 -\frac{1}{2} D_{\mu}\Phi^{\dagger} D_{\mu}\Phi -{\lambda \over 8} \,
 (\Phi^{\dagger}\Phi - f^2)^2 
\end{equation}
\begin{equation}
L_{S}  = - \frac{1}{2}\,\partial_{\mu} \vec S \, 
\partial_{\mu}\vec S
     -\frac{1}{2} m_{S}^2 \,\vec S^2 - \frac{\lambda_S}{8N} \, 
     (\vec S^2 )^2 
\end{equation}
\begin{equation}
L_{Interaction} = -\frac{\omega}{4\sqrt{N}}\, \, \vec S^2 \,\Phi^{\dagger}\Phi 
\end{equation}  

As before $\Phi=(\sigma+f+i\pi_1,\pi_2+i\pi_3)$ is the complex Higgs doublet of 
the SM with the
vacuum expectation value $<0|\Phi|0> = (f,0)$, $f=246$ GeV. Here, 
$\sigma$ is the physical  
Higgs boson and $\pi_{i=1,2,3}$ are the three Goldstone bosons. 
$\vec S = (S_1,\dots,S_N)$ is a real vector with 
$<0|\vec S|0>= \vec 0$. 
We consider this case, where the $O(N)$ symmetry stays unbroken,
because we want to concentrate on the effects of a finite width
of the Higgs particle. Breaking the $O(N)$ symmetry would lead
to more than one Higgs particle, through mixing. After the
spontaneous breaking of the SM gauge symmetry the $\pi$ fields
become the longitudinal polarizations of the vector bosons.
In the unitary gauge one can simply put them to zero. One
is then left with an additional interaction in the Lagrangian
of the form: 

\begin{equation}
L_{Interaction} = -\frac{\omega f}{2\sqrt{N}}\, \, \vec S^2 \,\sigma 
\end{equation}  
This interaction leads to a decay into the $\vec S$ particles, that
do not couple to other fields of the SM Lagrangian. On has therefore
an invisible width:
\begin{equation}
\Gamma_{Higgs}(invisible) = \frac {\omega^2}{32 \pi}\, \, 
\frac {f^2}{m_{Higgs}} 
(1-4 m_S^2/m_{Higgs}^2)^{1/2}
\end{equation}
This width is larger than the SM width even for moderate values of $\omega$,
because the SM width is strongly suppresed by the Yukawa coupings of the
fermions. Moreover one cannot exclude a large value of $\omega$, which 
would lead to a wide invisible Higgs. Limits on this model have been
put by the LEP collaborations in [17-19].
 \\

The general situation is thus, that after introducing singlets in the SM
the Higgs boson becomes of arbitrary shape and can have an arbitrary
branching ratio into invisible particles, where the invisible branching
ratio depends on the invariant mass point in the Higgs propagator. The only thing that
stays constant is the relative branching ratios into the SM decay products.
Direct limits on these type of models which I call singlet standard models (SSM) can therefore
be derived only on the basis of a decay mode independent analysis of the Higgs
signal. This analysis has been performed by the OPAL collaboration~\cite{opal}. From the analysis
one can conclude, that the K\"all\'en-Lehmann density integral $\int ds \, \rho(s)$
is small below roughly $\sqrt s < 90 \, GeV$,
the exact limits depending on the detailed model.
 From the SM limit $m_H < 190 \, GeV$ \cite{eww} one can also conclude
that the contribution $\int ds \, \rho(s)$ is small 
for $\sqrt s > 190 \, GeV$.

So far in this discussion I have shown that the SSMs are  possible and leave open
a  fairly large array of possibilities, with relatively small restrictions.
The SSM's are actually essentially undistinguishable from the SM with present data.
The natural question, which is often asked is therefore what is this good for?
Here cosmology appears to indicate that scalars indeed play a role in nature.
There are two important cosmological facts, that are relevant here, namely inflation and
the presence of dark matter. Recent results from the EGRET experiment can be 
interpreted as providing evidence of the annihilation of dark matter pairs \cite{deboer}.
For this mechanism to work cosmologically it is necessary that the dark matter
particles are stable by themselves through some preserved symmetry. In typical 
examples one takes here the R-parity in the MSSM and assumes the dark matter 
particles to be the LSP. However the above-mentioned $O(N)$ symmetry would 
do as well [22,23].
For inflation one typically takes a singlet field, that has no particular symmetry.
On this basis the "new minimal standard model" was introduced \cite{dav}. This model
contains an inflaton field $H$, a scalar field $S$, with a $Z_2$ symmetry
and singlet right-handed neutrinos $N_{\alpha}$ to allow for leptogenesis.  
The part of the Lagrangian containing the singlet scalar fields is then given
 by  the form:
\begin{equation}
L  = L_S + L_{H} + L_{S H N}
\end{equation}
where all interactions consistent with the symmetry are written down:
\begin{equation}
 L_S = -\frac{1}{2}\partial_{\mu}S\partial_{\mu}S -\frac{1}{2}m_S^2 S^2
   -g_S S^2 \Phi^{\dagger}\Phi -\lambda_S S^4
\end{equation}
\begin{equation}
L_{H} = -\frac{1}{2}\partial_{\mu}H\partial_{\mu}H -\frac{1}{2}m_H^2 H^2
   -\mu_H H^3 -\lambda_{H} H^4
\end{equation}
\begin{equation}
L_{S H N} = -\mu_{\Phi}H \Phi^{\dagger}\Phi -\mu_S H S^2 - \kappa_H H^2 \Phi^{\dagger}\Phi
   -\kappa_S H^2 S^2- (y_N^{\alpha\beta}H N_{\alpha}N_{\beta} + c.c.)
\end{equation}

This Lagrangian therefore contains a large number of extra parameters in particular Yukawa
couplings to the right-handed neutrino's and a number of singlet interactions.
From the previous discussion it is clear that these are not all necessary for renormalizability.
In particular one can put:
\begin{equation}
y_N^{\alpha\beta} = \kappa_S = \kappa_H = \mu_H = \lambda_{H} = 0
\end{equation}
If the $H$ field is higher dimensional these conditions are indeed also
necessary for renormalizability.
Since inflation appears to be driven by the mass term of the inflaton 
field \cite{astro}
this is consistent with the data and would explain why
 $\lambda_{H} < 10^{-14}$ \cite{dav}.
So this would mean that the minimal  model to explain also the
cosmological data would have only six parameters beyond the SM, where we have not
counted the neutrino masses. However there are two hidden discrete variables,
namely the dimension of the inflaton field and the $N$ of the $O(N)$ symmetry,
which was taken to be $Z_2$. 
In discussions the question often arises, whether putting the parameters in 
formula~[20] equal to zero, should be considered an unnatural fine-tuning, since
this relation is not due to a symmetry.
It must be emphasized, that the conditions are preserved under renormalization.
Therefore one cannot say that these conditions are completely arbitrary,
 as long as one has no deeper
insight in some form of underlying dynamics. The question is somewhat philosophical:
should one only consider the terms that are\,\, {\it needed for}\,\, renormalizability or assume
all terms to be present that are\,\, {\it allowed by}\,\, renormalizability? 
In most ordinary models these classes are the same, so the question does not arise.
 When the inflaton is 
higher-dimensional the situation is somewhat different. In this case renormalizability
forces one to the absence of inflaton self-couplings. One could therefore 
somewhat boldly turn the argument around and conclude that the experimental absence of
inflaton self-couplings is an indication, that higher dimensions play a role at least
in the scalar singlet sector.\\

What are the actual consequences for experiment when this picture is correct?
As mentioned before the Higgs mass density will be concentrated in the range
$90\, GeV< \sqrt s < 190\, GeV$ on the basis of the LEP data and other precision
electroweak data. This makes this type of Higgs boson extremely hard to see,
paricularly in the general SSM. There are essentially four scenarios.

The first is the optimistic one. The Higgs particle splits into a number of peaks,
let us say two with an equal weight, one decaying in SM particles,
the other invisible. Since already a fully invisibly decaying Higgs boson
is marginal, this peak will indeed not be seen. What one therefore sees is
a SM Higgs peak, with a reduced production cross section, here one half of the
SM. However it could also be for instance 20\% or 30\% or less or it could be 80\%.
In particular the last case shows, that it is absolutely necessary to know the
Higgs production cross section as precisely as possible: one must try to distinguish
between 80\% or 100\% of a Higgs boson. This scenario is ideal, because one 
would have established at the same time that the Higgs boson is there and also
that the SM is not complete, thereby providing an excellent argument for building
the ILC in order to find the dark matter of the universe.
 
The second one is the moderately pessimistic/optimistic one of a completely
invisibly decaying Higgs boson. Recent analyses have shown that a  reasonable
signal is possible in this case, 95\% CL contours are given in \cite{jakobs}.
 However these analyses
have only been performed for a narrow single Higgs. This is maybe not crucial,
since there is anyway no mass peak in the missing energy signal. However how
serious this is can only be determined by a detailed experimental analysis.
In this case the ILC can clarify the situation without difficulty.

The third one is the moderately pessimistic/optimistic scenario of a completely
spread out Higgs boson decaying with SM branching ratios. This could be quite
difficult if one is unlucky, depending on where the K\"all\'en-Lehmann weight lies.
 For instance
60\% spread out flatly  over the range $100\, GeV < \sqrt s < 130\, GeV$ and 40\%
around $160\, GeV$ would be a real experimental challenge. This possibility 
emphasizes the necessity of correlated searches in different channels,
which would require further refinements in present day analysis tools.
This situation can also be studied at the  ILC without difficulty.
The analysis in \cite{beauch}, considering a six-dimensional
singlet gives a too optimistic view of the situation,
as it has ignored the singlet-doublet mixing in the model, which is the
dominant effect. 

The fourth possibility, which at the moment appears to be consistent with the 
cosmological data is that actually the model is right, but that the singlets
are too heavy for the Higgs boson to decay into. In this case one sees only
the SM at the LHC and nothing else. Only a much more precise understanding
of cosmological limits and dark matter characteristics, possibly by direct
detection could give a convincing argument to pursue the ILC in this case.\\

{\bf Acknowledgement} I thank Dr. A. Ferroglia, Prof. M. Veltman and
Prof. K. Jakobs for  critical readings of the 
manuscript.\\

\end{document}